\documentclass[preprint,floatfix,onecolumn,superscriptaddress]{revtex4-1}
\usepackage[dvips]{graphicx}
\usepackage[english]{babel}
\usepackage{amsmath}
\usepackage{amssymb}
\usepackage{tensor}
\usepackage{subfigure}

\newcommand{\be}{\begin{eqnarray}}
\newcommand{\ee}{\end{eqnarray}}

\newcommand{\dc}{c^{\dagger}}

\begin{document}

\title{Mobility Edges in one-dimensional Models with quasi-periodic disorder}

\author{Qiyun Tang}
\affiliation{College of Physics, Sichuan University, Chengdu, Sichuan 610064, China}

\author{Yan He}
\affiliation{College of Physics, Sichuan University, Chengdu, Sichuan 610064, China}
\email{heyan$_$ctp@scu.edu.cn}

\begin{abstract}
We study the mobility edges in a variety of one-dimensional tight binding models with slowly varying quasi-periodic disorders. It is found that the quasi-periodic disordered models can be approximated by an ensemble of periodic models. The mobility edges can be determined by the overlaps of the energy bands of these periodic models. We demonstrate that this method provides an efficient way to find out the precise location of mobility edge in qusi-periodic disordered models. Based on this approximate method, we also propose an index to indicate the degree of localization of each eigenstate.

\end{abstract}

\maketitle

\section{Introduction}

More than fifty years ago, P. W. Anderson pointed out in a seminal work \cite{Anderson} that the disordered on-site potential will make the electronic wavefunction become localized at certain site rather than propagating around the whole system like a plane wave. Since then, the Anderson localization was extensively studied in the condensed matter physics. In three-dimensional (3D) systems, the Anderson localization will occur only when the disorder strength is larger than certain value. When the disorder is not strong enough, the band structure will contain both extended or localized eigenstates. This leads to the concept of mobility edge which is the energy separating the localized and extended energy levels\cite{Mott}. The scaling theory points out that the dimension is an important factor for Anderson localization, and the critical point for the existence of mobility edge is 2D \cite{scaling1,scaling2}, that is, in 2D and 1D systems, even if the disorder intensity is infinitesimal small, the mobility edge cannot appear, and all wave functions are localized.

For a very long time, physicist only focused on the study of systems with uncorrelated disorders. Stating from 80's, low-dimensional quasi-periodic systems have been widely investigated. Due to the highly correlated disorders, it is still possible to have mobility edges even in low-dimensional systems\cite{1D1,1D2,1D3,1D4,1D5,1D6,1D7}. A famous example is the so called Aubry-Andre (AA) model \cite{Aubry,Harper} which is simply a 1D tight-binding model with incommensurate on-site potentials. One can show that the AA model has self-dual symmetry. With the increasing of the incommensurate on-site potentials, all the eigenstates of the AA model will undergo the transition from extended to localized, and the self-dual point is the transition point between the localized and extended states. Therefore, there is no mobility edge in the standard AA model, but if the model is further modified, such as the introduction of long-range hopping, then the mobility edge can appear in the improved AA model\cite{hopping1,hopping2,hopping3,hopping4,hopping5,hopping6,hopping7}.

Through the investigation of the AA model and its variants, a new class of system with slowly varying incommensurate potential has been extensively studied by Das Sarma and his co-workers \cite{Xie1,Xie2}. They found two mobility edges and proposed a heuristic semi-analytical method to solve the mobility edge in such systems. This further stimulated a lot of more theoretical works on the mobility edges in 1D quasi-periodic systems \cite{Sun,Huse,Biddle}, including generalizing the disorders to the off-diagonal hopping terms \cite{Liu1} or to some topological 1D model such as Kitaev chain \cite{Liu2}. Among them, the off-diagonal hopping AA model has attracted extensive research interests, because it contains rich and novel physical phenomena such as zero-energy topological edge states\cite{non1,non2}.

In the present paper, we will study the generalized AA model with slow varying quasi-periodic disorders in hopping and potential terms. We will first introduce the heuristic semi-analytical method to determine the mobility edges. This method is based the property that the slow varying disorder approach to a constant asymptotically. Although this method is very useful for some simple models, it also has severe limitation in finding mobility edges for models with more complicated disorders. To overcome this difficulty, we approximate the quasi-disordered model with an emsemble of different periodic models. Then the region of extended states can be approximately obtained by the overlaps of energy bands of these periodic models. We will demonstrate this method in details for several variations of AA model and also the Su-Schrieffer-Heeger (SSH) model \cite{SSH1}. Based on this method, we also propose a new index to indicate the degree of localization of eigenvectors.

\section{A semi-analytical method to determine the mobility edge}

We consider the AA model with slowly varying quasi-periodic disorders.
\be
H=-\sum_{i=1}^{N-1}(t+w_i)(\dc_i c_{i+1}+\dc_{i+1} c_i)+\sum_{i=1}^N\mu_i \dc_i c_i
\label{AAH}
\ee
Here $N$ is the total number of lattice sites. For convenience, we assume that $t = 1$ as the energy unit. The slowly varying disorder is given by
\be
w_i=w\cos(2\pi\alpha i^v),\quad
\mu_i=\mu\cos(2\pi\beta i^v+\phi)
\label{w}
\ee
Here $w$ and $\mu$ is the amplitude of the disorder. $\alpha$ and $\beta$ are certain irrational numbers, which determines the quasi-period. For example, one can assume that $\alpha=\beta=(\sqrt{5}-1)/2$. Actually, whether $\alpha$ is irrational or rational does not affect the results. This is because $0<v<1$ in Eq.(\ref{w}), which means that $i^{v}$ is irrational number and make the system quasi-periodic. For later convenience, we will assume that $\alpha=\beta=0.5$. Here $\phi$ is certain possible initial phase angle.

Since the exponent $v$ satisfies  $0<v<1$, we have
\be
\frac{d w_i}{d i}=-2wv\pi\alpha i^{v-1}\sin(2\pi\alpha i^v),\qquad
\lim_{i\to\infty}\frac{d w_i}{d i}=0
\label{dw}
\ee
Therefore, because the power $0<v<1$, the disorder term eventually become a constant for large enough $i$.

First, we discuss the semi-analytical method to determine the energy region of extended states or the locations of mobility edges. As an example, we first consider a simple case where $\beta=0$, thus the potential is just a constant $\mu$. If we assume that the eigenstate is $\psi=(a_1,\cdots,a_N)^T$, then the eigenvector equation in the real space is given by
\be
(t+w_j)a_{j-1}+(E-\mu)a_j+(t+w_j)a_{j+1}=0
\ee
Since $w_j$ approaches to a constant as $j$ become very large, we find the following asymptotic eigenvector equation
\be
(t+w_{\infty})a_{j-1}+(E-\mu)a_j+(t+w_{\infty})a_{j+1}=0
\ee
Here $w_{\infty}=\lim_{j\to\infty}w_j$. As usual, the eigenvector is assumed to take the form $a_j\propto\lambda^j$, then we arrive at the following characteristic equation
\be
\lambda^2+\frac{E-\mu}{t+w_{\infty}}\lambda+1=0
\ee
which gives two roots
\be
\lambda_{1,2}=\frac{-A\mp\sqrt{A^2-4}}{2},\quad A=\frac{E-\mu}{t+w_{\infty}}
\ee
In order to find extended states, $\lambda_{1,2}$ must be complex numbers with unit modular $|\lambda_{1,2}|=1$. Therefore, the condition for extended states is
\be
\Big|\frac{E-\mu}{t+w_{\infty}}\Big|<2
\label{ineq}
\ee
Since $w<t$, the most stringent inequality one can get from Eq.(\ref{ineq}) is obtained by taking $w_{\infty}=-w$. Therefore,we find the following condition for extended states
\be
\mu-2(t-w)<E<\mu+2(t-w)
\ee
which agrees with the mobility edge shown in Figure 1 (a).

In Figure 1, the localization of the $n$-th eigen-wavefunction $\psi_n$ is indicated by the inverse participation ratio (IPR) \cite{Thouless,Kohmoto,Schreiber}, which is defined as
\be
\mbox{IPR}_n=\sum_{j=1}^N\Big|a^n_j\Big|^4,\quad
\psi_n=(a^n_1,\cdots,a^n_N)
\ee
Here we assume that $\psi_n$ is normalized, i.e $\sum_{j=1}^N\Big|a^n_j\Big|^2=1$. For the extend states, one expects that each component of $\psi_n$ is roughly the same order of magnitude $|a^n_j|^2\sim 1/N$ for all $j$. Therefore, we find that IPR$\sim \sum_{j=1}^N\frac{1}{N^2}=\frac1N$, which is very small for the extended states. One the other hand, the non-zero amplitude of localized states will mostly confined to only a few components, therefore in this case, we expect that the IPR will be order 1.

In Figure \ref{fig-AAH} (a), we plot the eigenenergy of AA mdoel of Eq. (\ref{AAH}) as a function of $\mu$. The parameters used in this calculation is listed in the figure caption. The color of each data points represents the IPR value of the corresponding eigenstates. The brighter colors indicate larger values of IPR which represents a localized states. The darker colors indicate that IPR is close to zero which represents extended states. One can see that the dark colored area is precisely confined between the red and green solid lines which are the mobility edges we have obtained from the semi-analytic calculations.  In order to make a more quantitatively observation, in Figure \ref{fig-AAH} (c), we also plot the IPR as a function of eigenenrgy for a few selected points $\mu=0,\,0.5,\,5$. One can see that the IPR drops from the order of magnitude of $10^{-2}$ to $10^{-4}$ when the eigenenergy across some critical values. This clearly signals the transition from localized states to extended states. These critical values precisely matches the mobility edges we have discussed.

We can also use the same method to consider a more complicated case like $\beta=\alpha$, i.e. both $\mu$ and $w$ are quasi-periodic disordered. Following similar steps, we find the condition for extended states is given by
\be
\Big|\frac{E-\mu_{\infty}}{t+w_{\infty}}\Big|<2
\ee
Unfortunately, it is difficult to determine the precise boundary between extended and localized states from the above inequality. In the next section, we will introduce a new approximate method to determine the location of mobility edges.

\section{An alternative method to determine the mobility edge}

\begin{figure}
  \centering
   \subfigure[]{
  		\begin{minipage}[t]{0.45\linewidth}
  			\centering
  			\includegraphics[width=\textwidth]{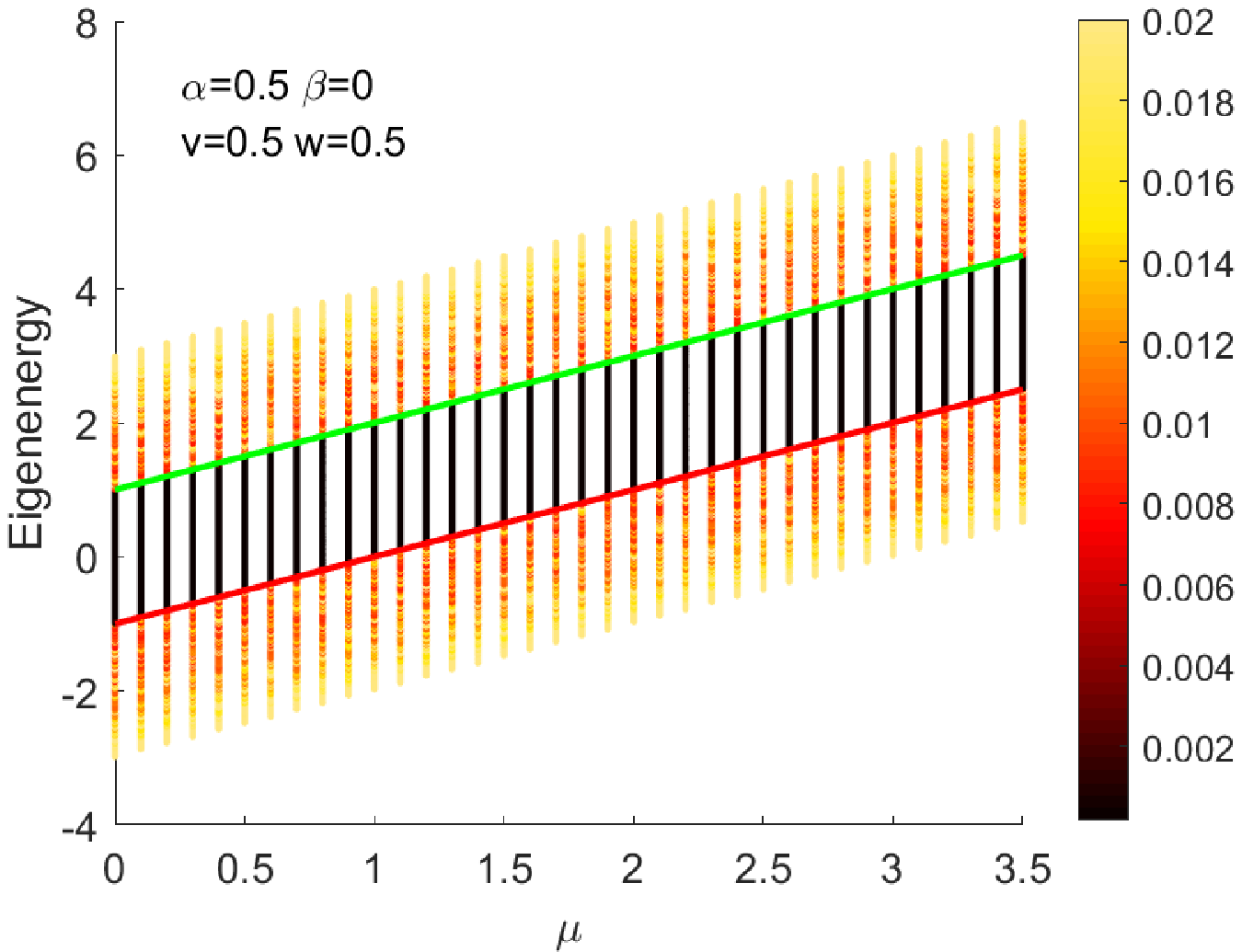}
  		\end{minipage}
  	}
  	\subfigure[]{
  		\begin{minipage}[t]{0.45\linewidth}
  			\centering
  			\includegraphics[width=\textwidth]{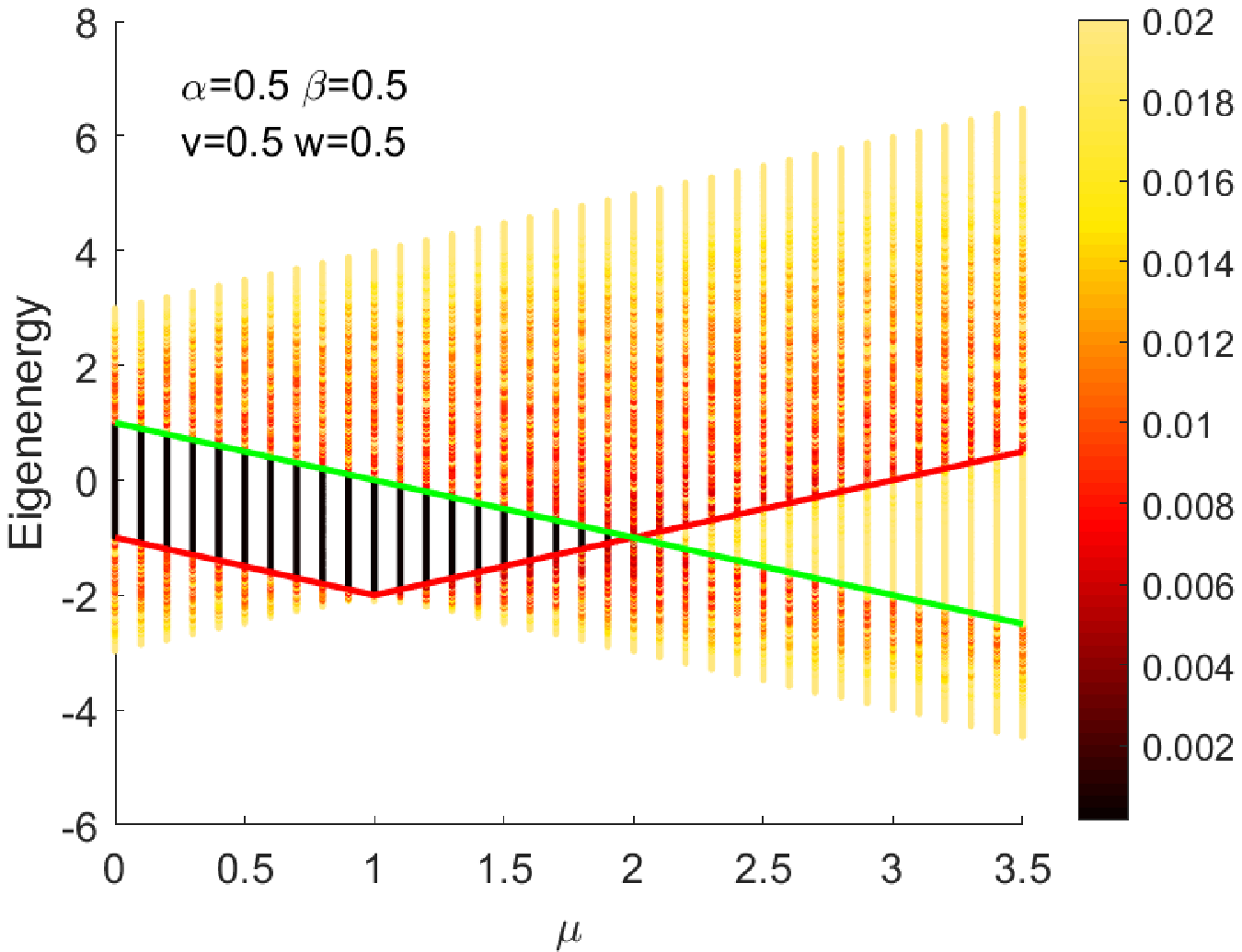}
  		\end{minipage}
  	}
  	\quad
  	\subfigure[]{
  		\begin{minipage}[t]{0.45\linewidth}
  			\centering
  			\includegraphics[width=\textwidth]{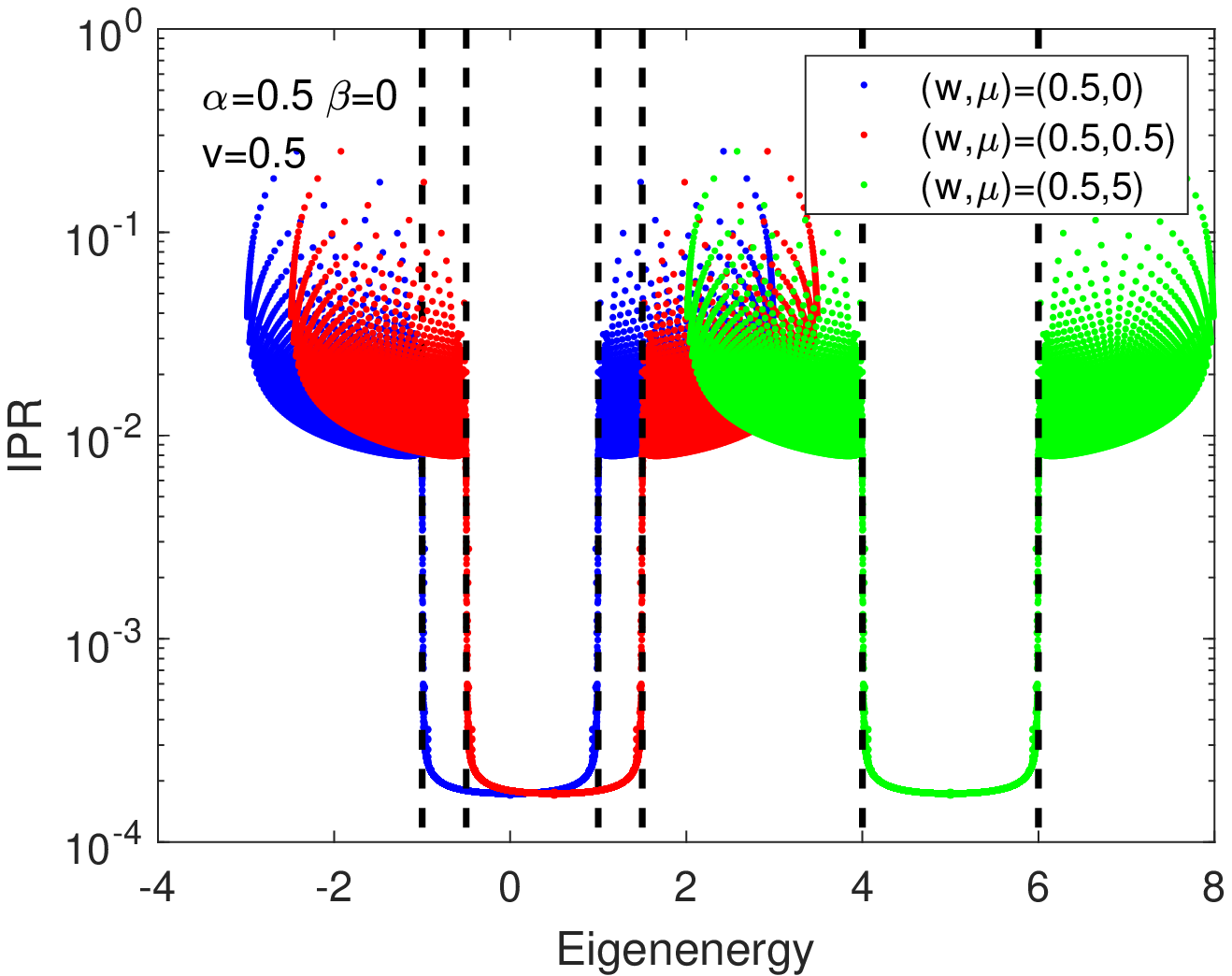}
  		\end{minipage}
  	}
  	\subfigure[]{
  		\begin{minipage}[t]{0.45\linewidth}
  			\centering
  			\includegraphics[width=\textwidth]{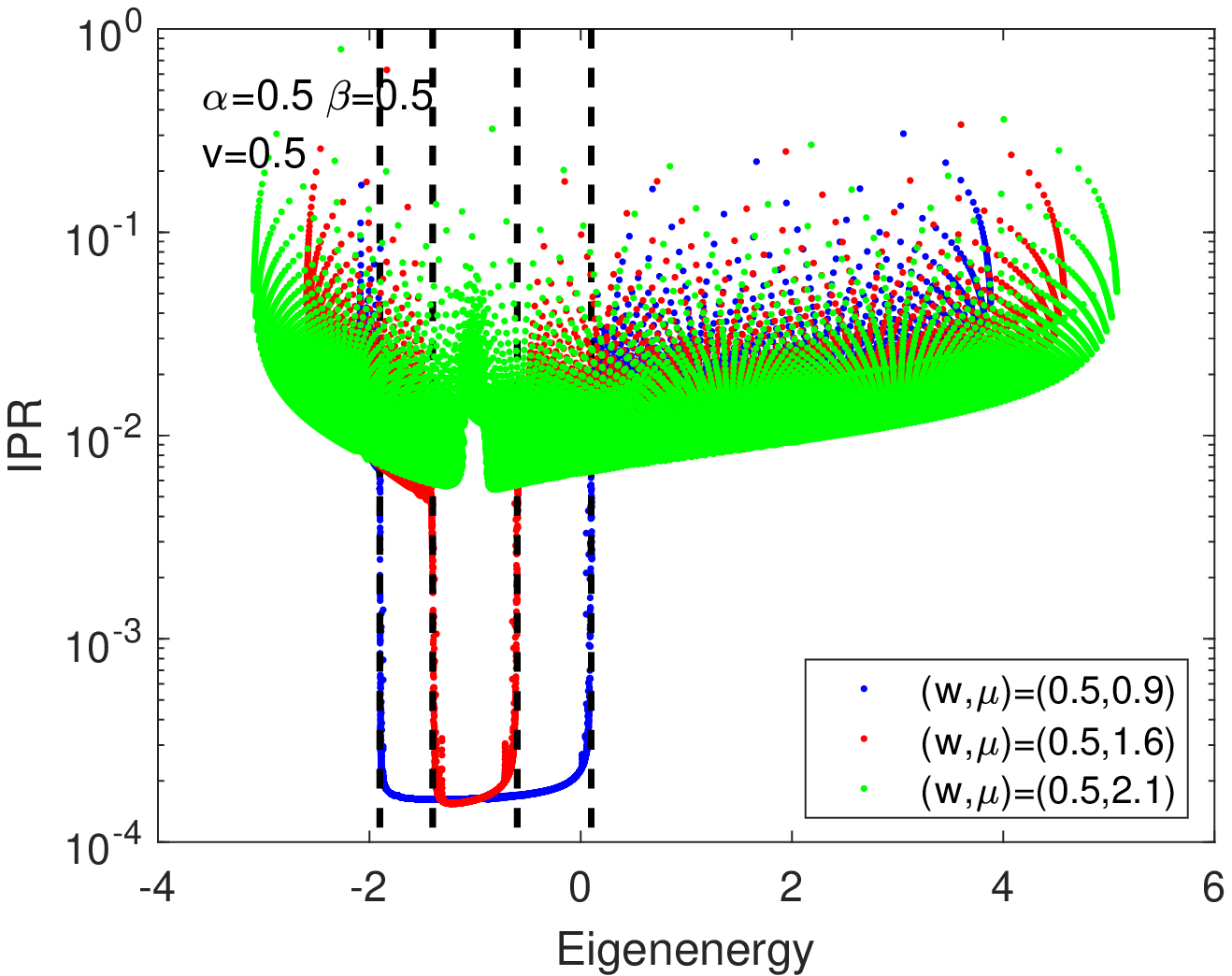}
  		\end{minipage}
  	}
  	\centering
\caption{(a) Eigenenergy of AA model Eq. (\ref{AAH}) as a function of $\mu$ with $\alpha=0.5$, $\beta=0$, $w=0.5$ and $v=0.5$.  (b) Eigenenergy of AA model Eq. (\ref{AAH}) as a function of $mu$ with $\alpha=0.5$, $\beta=0.5$, $w=0.5$ and $v=0.5$. The colors in panels (a) and (b) represent the IPR of each eigenstates. The red and green curves represent the mobility edges. (c) The IPR as a function of eigenenergy for $(w,\mu)=(0.5,0),\,(0.5,0.5),\,(0.5,5)$, other parameters are the same as (a). (c) The IPR as a function of eigenenergy for $(w,\mu)=(0.5,0.9),\,(0.5,1.6),\,(0.5,2.1)$, other parameters are the same as (b).}
\label{fig-AAH}
  \end{figure}

Now we turn to discuss a new method, which can determine the extended eigenstates more efficiently. The basic idea of this method is to approximate the above quasi-periodic disordered model by an ensemble of periodic models which can be denoted as $M_a$. The Hamiltonian of $M_a$ is given by
\be
H_a=-\sum_{i=1}^{N}\Big[(t+w_a)(\dc_i c_{i+1}+\dc_{i+1} c_i)+\mu_a \dc_i c_i\Big],\quad
a=1,\cdots N
\ee
Here we have assumed periodic boundary condition such that $c_{N+1}=c_1$. For each model $M_a$, since $w_a$ and $\mu_a$ are constant, we can diagonalized the above Hamiltonian in momentum space and find extended Bloch wave function with the following energy band
\be
E=\mu_a-2(t+w_a)\cos k
\ee
Therefore, the region of extended states for the model $M_a$ is just the range of the energy band
\be
\mu_a-2(t+w_a)<E<\mu_a+2(t+w_a)
\ee
One can imagine that for the quasi-periodic disordered model, the extended states will appear for a given energy $E$, if this $E$ is inside the energy bands of all the above models $M_a$. Therefore, we expect that the energy region of extended state of Eq.(\ref{AAH}) is given by
\be
E\in \bigcap_a\Big(\mu_a-2(t+w_a),\,\mu_a+2(t+w_a)\Big)
\label{E_a}
\ee
We can roughly call the above method as ``energy matching method''. Now we can apply this formula to determine the mobility edge of AA model in the following 4 different situations.

\textbf{AA model with disordered hopping}

For the case of $\beta=0$, $\mu_a$ is just a constant $\mu$. Then the intersection of all the above intervals is just the smallest possible interval for certain $a$.
It is easy to see that the smallest interval corresponds to $w_a=-w$. Therefore, we find that the energy region of extended state is
\be
E\in \Big(\mu-2(t-w),\,\mu+2(t-w)\Big)
\ee
which recovers the results we have obtained in the last section.

\textbf{AA model with the same disorders in hopping and potential}

Now we turn to a more complicated model with $\beta=\alpha$. In other word, $\mu_a$ and $w_a$ have the same disorder. We have seen that the semi-analytic method is not very useful in this case. We will see that our energy matching method provides us a better way to determine the mobility edges. For a given $\mu$, the smallest interval is again given by taking $w_a=-w$. Since $\mu_a=\frac{\mu}{w}w_a$, we also have $\mu_a=-\mu$ at the same time. Therefore, the energy region of extended state should be given by
\be
E\in \Big(-\mu-2(t-w),\,-\mu+2(t-w)\Big)
\label{E1}
\ee
The other extreme values of $w_a=w$ and $\mu_a=\mu$ will give us another lower bound $E_{min}=\mu-2(t+w)$. When $\mu<2w$, this lower bound is too low comparing with the one from Eq.(\ref{E1}). But when $\mu>2w$, this lower bound will be higher than the previous one, which become a new lower bound. In summary, when $\beta=\alpha$, we find the condition for extended states is given by
\be
&&E\in\Big(-\mu-2(t-w),\,-\mu+2(t-w)\Big),\quad \hbox{for}\, 0<\mu<2w\nonumber\\
&&E\in\Big(\mu-2(t+w),\,-\mu+2(t-w)\Big),\quad \hbox{for}\,  \mu>2w
\label{edge}
\ee
To verify the above mobility edges, we plot the eigenenergy of AA mdoel with disorders in both hopping and potential in Figure \ref{fig-AAH} (b). The parameters used in this calculation is listed in the figure caption. Again, the color of each points represents the IPR value of the eigenstates. One can see that the dark area matches the mobility edges we have decided above by the energy matching method. An interesting feature of this model is that the lower mobility edges is not monotonic. This trend is not easy to get in the traditional method. In Figure \ref{fig-AAH} (d), the IPR values as a function of eigenenergy is shown for a few points $\mu=0.9,\,1.6,\,2.1$. It provides a more quantitative evidence of the mobility edges of Eq.(\ref{edge}).

\begin{figure}
  \centering
   \subfigure[]{
  		\begin{minipage}[t]{0.45\linewidth}
  			\centering
  			\includegraphics[width=\textwidth]{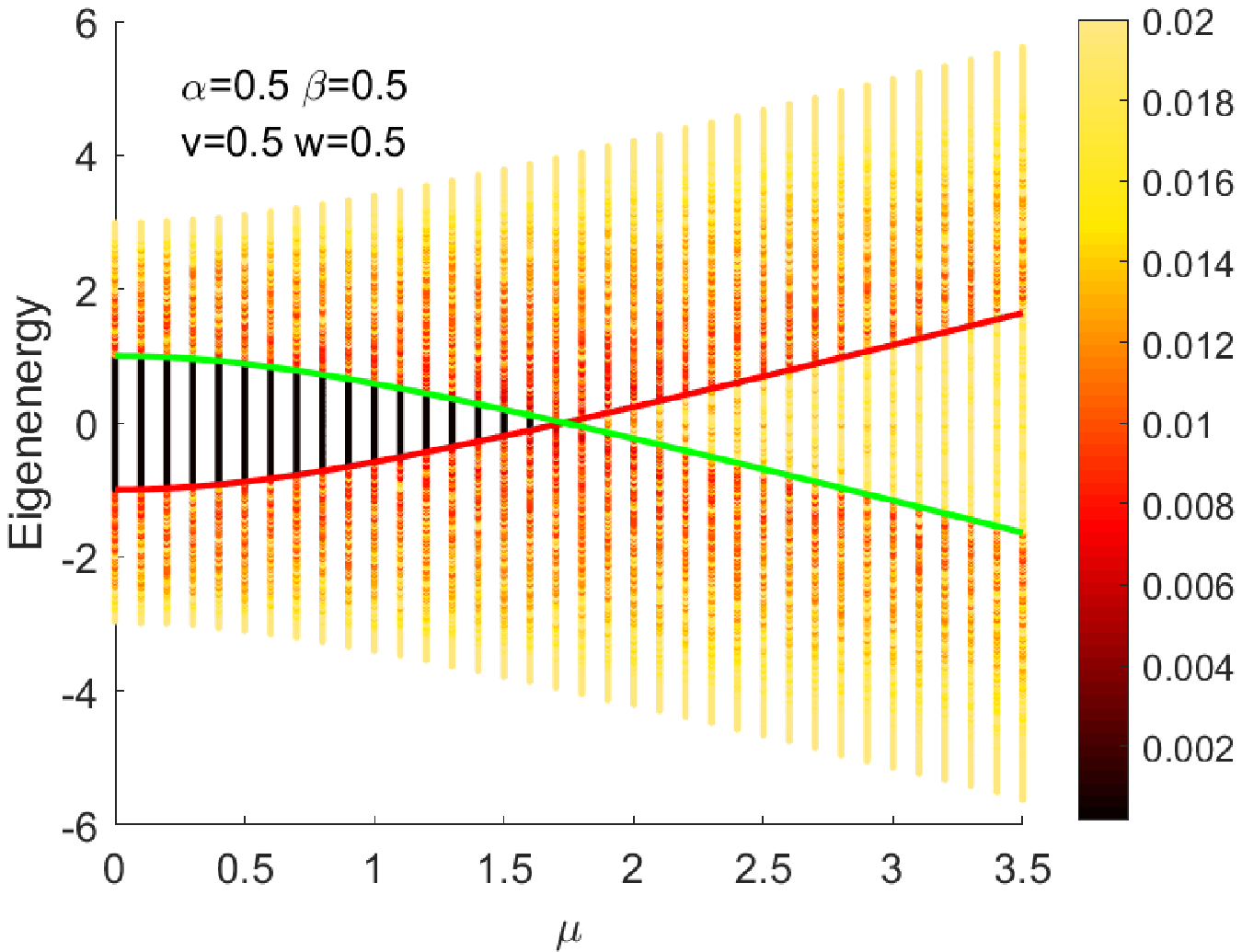}
  		\end{minipage}
  	}
  	\subfigure[]{
  		\begin{minipage}[t]{0.45\linewidth}
  			\centering
  			\includegraphics[width=\textwidth]{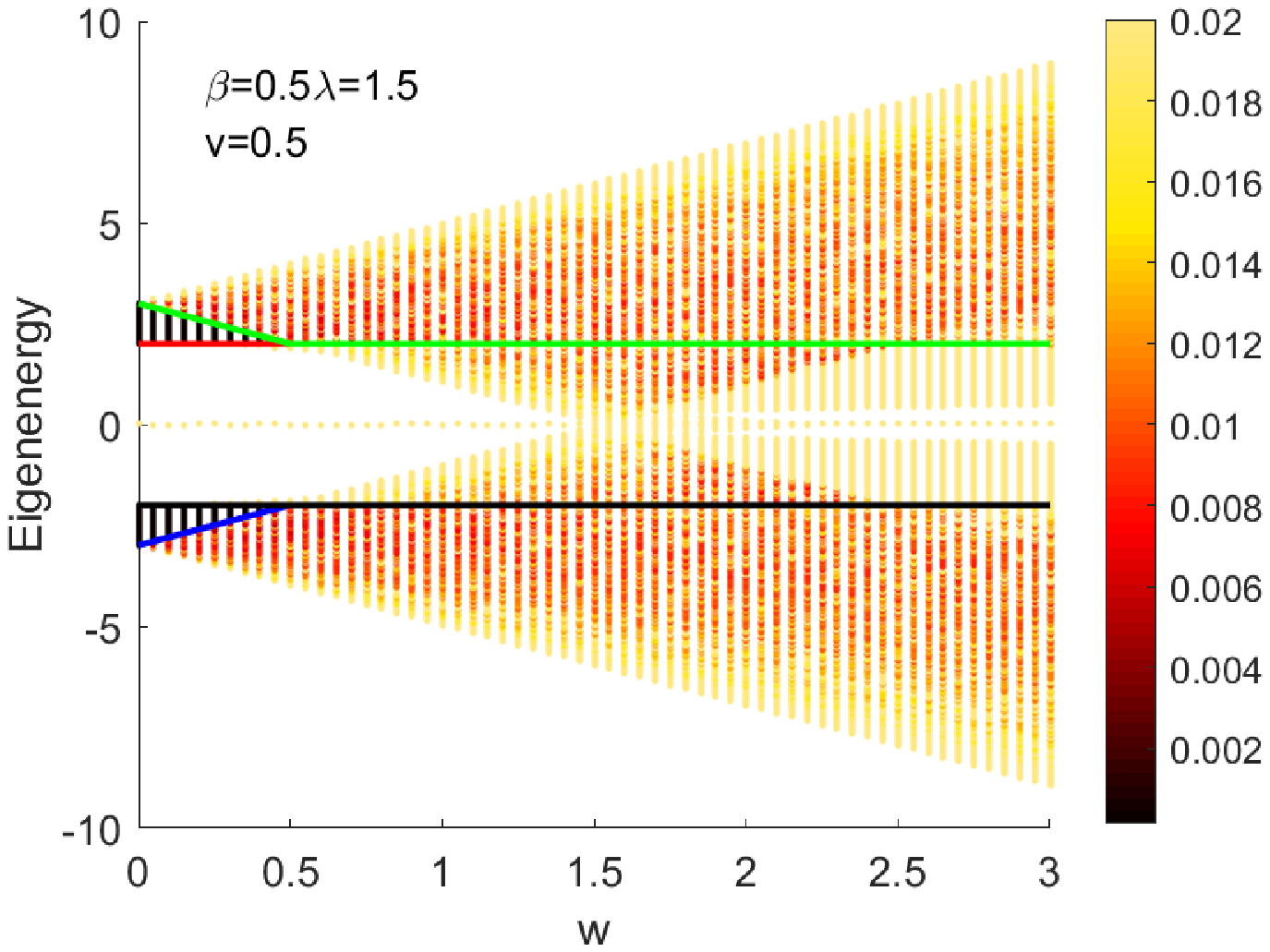}
  		\end{minipage}
  	}
  	\quad
  	\subfigure[]{
  		\begin{minipage}[t]{0.45\linewidth}
  			\centering
  			\includegraphics[width=\textwidth]{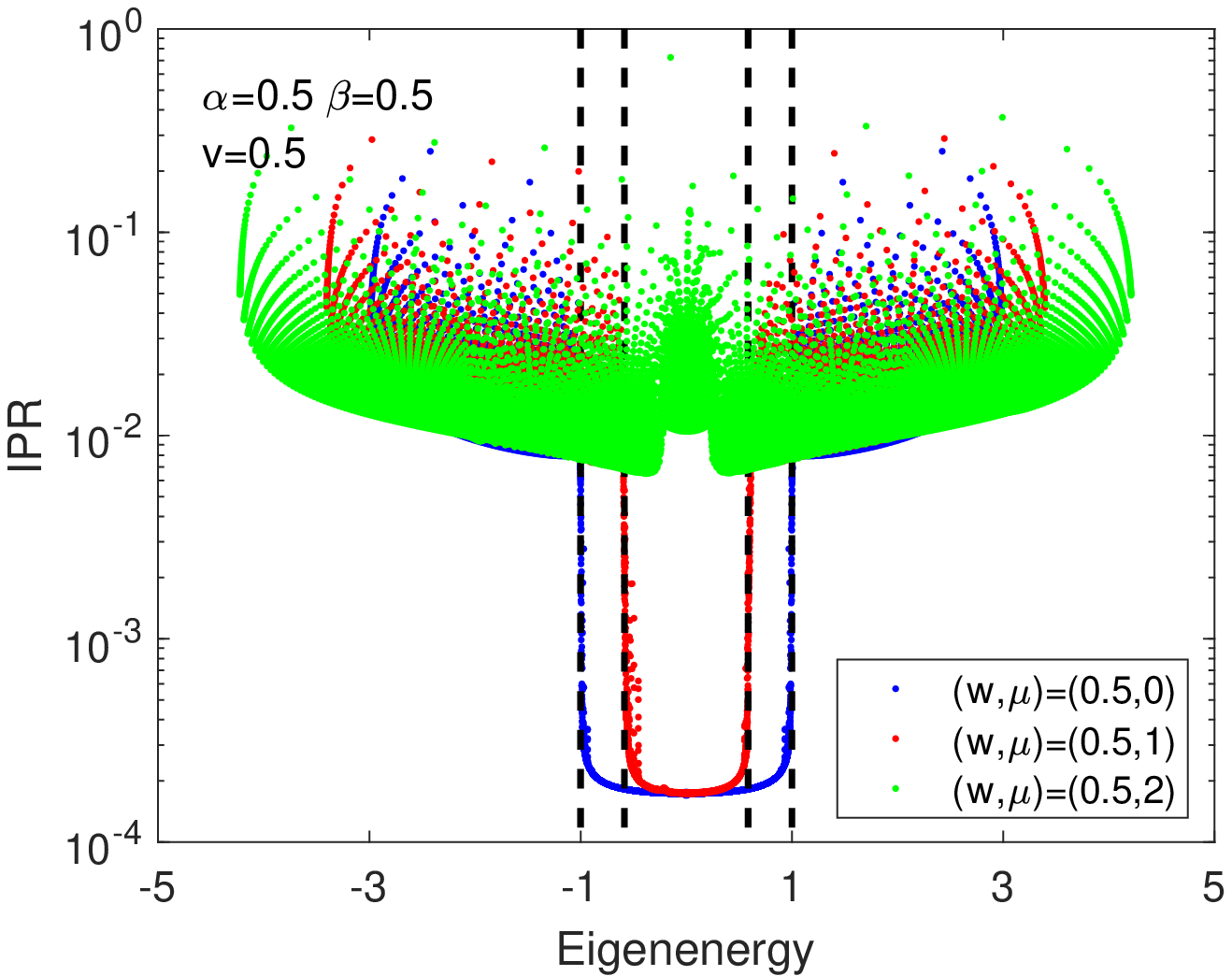}
  		\end{minipage}
  	}
  	\subfigure[]{
  		\begin{minipage}[t]{0.45\linewidth}
  			\centering
  			\includegraphics[width=\textwidth]{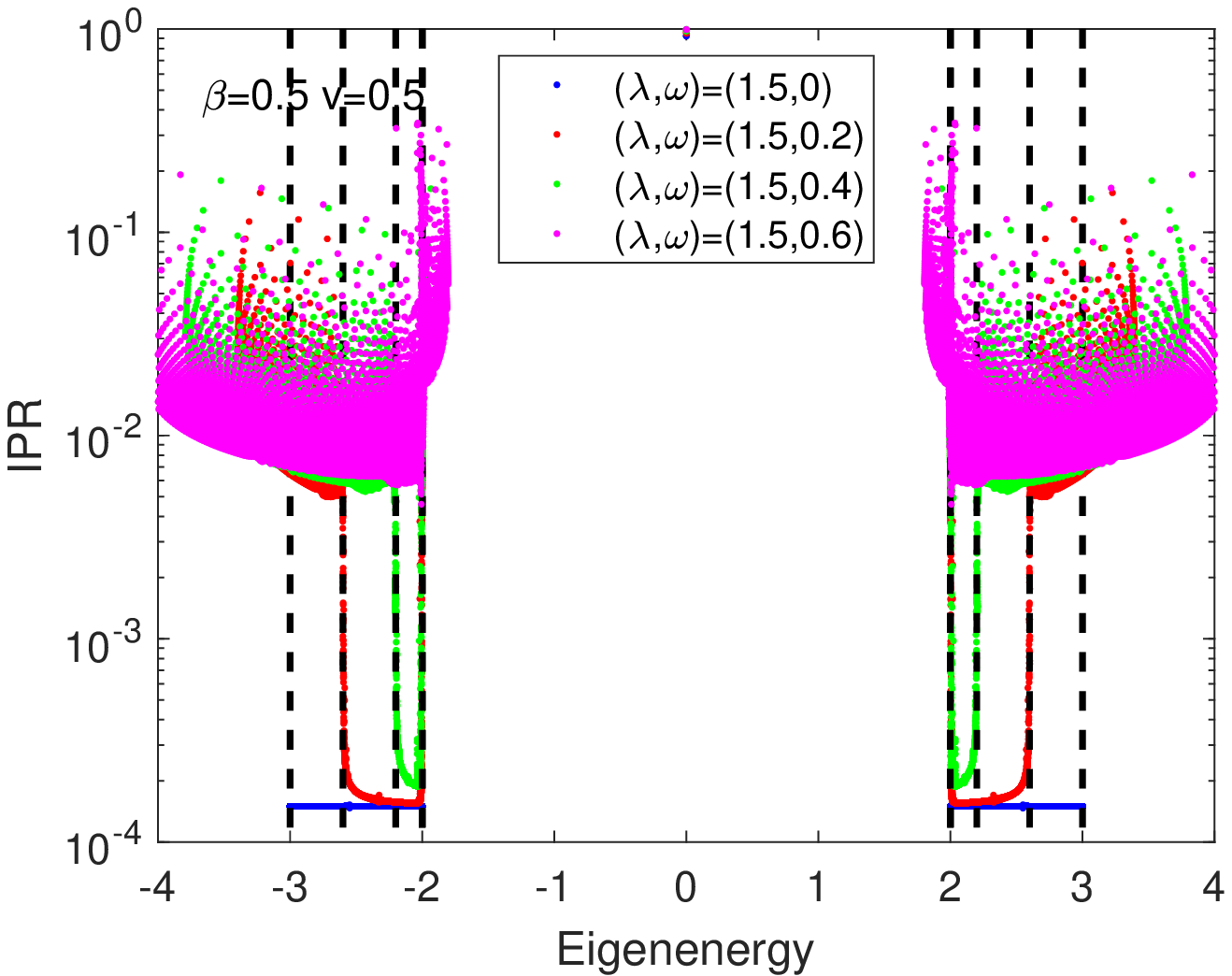}
  		\end{minipage}
  	}
  	\centering
\caption{(a) Eigenenergy of AA model Eq. (\ref{AAH}) as a function of $\mu$ with $\alpha=0.5$, $\beta=0.5$, $w=0.5$ and $v=0.5$, $\phi=\pi/2$.  (b) Eigenenergy of SSH model Eq. (\ref{SSH}) as a function of $w$ with $\lambda=1.5$, $\beta=0.5$ and $v=0.5$. The colors in panels (a) and (b) represent the IPR of each eigenstates. The red and green curves represent the mobility edges. (c) The IPR as a function of eigenenergy for $(w,\mu)=(0.5,0),\,(0.5,1),\,(0.5,2)$, other parameters are the same as (a). (c) The IPR as a function of eigenenergy for $(\lambda,\mu)=(1.5,0),\,(1.5,0.2),\,(1.5,0.4),\,(1.5,0.6)$, other parameters are the same as (b).  }
\label{fig-SSH}
  \end{figure}

\textbf{AA model with different disorders in hopping and potential}

As a more exotic example, we can have different disorders in hopping and potential. This can be easily achieved by setting $\phi\neq0$. In this case, $w_a$ and $\mu_a$ have relatively independent disorder, but they are not completely independent, and there is a phase difference between them. For the convenience of demonstration, here we choose $w_a=w\cos(2\pi\alpha a^v)$ and $\mu_a=\mu\cos(2\pi\beta a^v+\pi/2)=\mu\sin(2\pi\beta a^v)$, and the phase difference between them is $\phi=\dfrac{\pi}{2}$. Since $w_a$ and $\mu_a$ are no longer the same disorder, we use the auxiliary angle formula to further simplify the Eq.(\ref{E_a})
\be
E\in \bigcap_a\Big(R_{-a}-2t,\,R_{+a}+2t\Big)
\ee
Where $R_{\pm a}$ are given by
\be
R_{\pm a}=R\cos(\pm 2\pi\alpha a^v-\arctan(\dfrac{\mu}{2w})),\quad R=\sqrt{\mu^2+(2w)^2}
\ee
For a given $R$, the smallest interval is obtained by taking $R_{-a}=R$ for lower bound and $R_{+a}=-R$ for upper bound. Therefore, we simply find that the region of extended states is given by
\be
E\in \Big(R-2t,\,-R+2t\Big)\label{edge1}
\ee
To verify the above mobility edges, we plot the eigenenergy of this model as a function $\mu$ in Figure \ref{fig-SSH} (a). The IPR values are reflected by the colors of each points.  One can see that the dark area matches the mobility edges of Eq.(\ref{edge1}). In Figure \ref{fig-SSH} (c), we plot the IPR values as a function of eigenenergy for a few points $\mu=0,\,1,\,2$. It provides a more precise numerical evidence of the mobility edges of Eq.(\ref{edge1}).

\textbf{SSH model with disordered hopping}

In addition to AA model, SSH model is also an extensively studied 1D topological model \cite{SSH2,SSH3,SSH4,SSH5}. When the slow varying quasi periodic disorder is applied to the SSH model, mobility edge will also appear \cite{Liu3}. Because this kind of model is more comlicated than AA model, the semi analytical method is relatively difficult, and the advantages of the new method are more obvious. The Hamiltonian of SSH model is given by
\be
H=\sum_{i=1}^{N}(t_1-w_i)(\dc_{i,A} c_{i,B}+\dc_{i,B} c_{i,A})+\sum_{i=1}^{N-1}(t_2+w_i)(\dc_{i,B} c_{i+1,A}+\dc_{i+1,A} c_{i,B})
\label{SSH}
\ee
Here $w_i=w\cos(2\pi\alpha i^v)$ and assume that $t_{1,2}=t\pm \lambda$. We approximate this model by an ensemble of models without disorder as
\be
H_a=\sum_{i=1}^{N}(t_1-w_a)(\dc_{i,A} c_{i,B}+\dc_{i,B} c_{i,A})+\sum_{i=1}^{N-1}(t_2+w_a)(\dc_{i,B} c_{i+1,A}+\dc_{i+1,A} c_{i,B})
\ee
Here $a=1,\cdots,N$. For a fix $a$, then energy band is given by
\be
E=\pm\sqrt{(t_1-w_a)^2+(t_2+w_a)^2+2(t_1-w_a)(t_2+w_a)\cos k}
\ee
Since the energy band is symmetric about $E=0$, we can focus on the upper band, the lower one is the same. To determine the mobility edges, we have to distinguish the following two case.

(1) When $\lambda<t-w$, the energy band is located inside the following region $E\in \Big(2\lambda+2w_a,\,2t\Big)$. Therefore, for the disordered model, we find the condition of extended states is given by
\be
E\in \bigcap_a\Big(2\lambda+2w_a,\,2t\Big)
\ee
Clearly, the intersection is given by the smallest interval corresponding to $w_a=w$. Thus the condition of extended states for this case
is
\be
E\in\Big(2\lambda+2w,\,2t\Big)
\ee

(2) On the other hand, if $\lambda>t+w$, the energy band is located inside the following region $E\in \Big(2t,\,2u+2w_a\Big)$. Then, the condition of extended states for the disordered model is given by
\be
E\in \bigcap_a\Big(2t,\,2\lambda+2w_a\Big)
\ee
Then the intersection is given by the smallest interval corresponding to $w_a=-w$. Thus the condition of extended states for this case
is
\be
E\in\Big(2t,\,2\lambda-2w\Big)\label{edge-SSH}
\ee
To verify the mobility edges of this case, we plot the eigenenergy of the SSH model Eq.(\ref{SSH}) as a function $w$ in Figure \ref{fig-SSH} (b). The parameters used in this calculation is listed in the figure caption. The color of each point represents the IPR value of the eigenstates. The larger IPR values correspond to brighter colors. One can see that the dark area matches the mobility edges of Eq.(\ref{edge-SSH}). In Figure \ref{fig-SSH} (d), we also plot the IPR values as a function of eigenenergy for a few points $\mu=0.2,\,0.4,\,0.6$, which verify the mobility edges more precisely.

For completeness, we mention that there is no extended states for $t-w<\lambda<t+w$. This can be easily verified numerically (Not shown in the Figure).

\section{Discussion and an index of localization}

\begin{figure}
  \centering
   \subfigure[]{
  		\begin{minipage}[t]{0.45\linewidth}
  			\centering
  			\includegraphics[width=\textwidth]{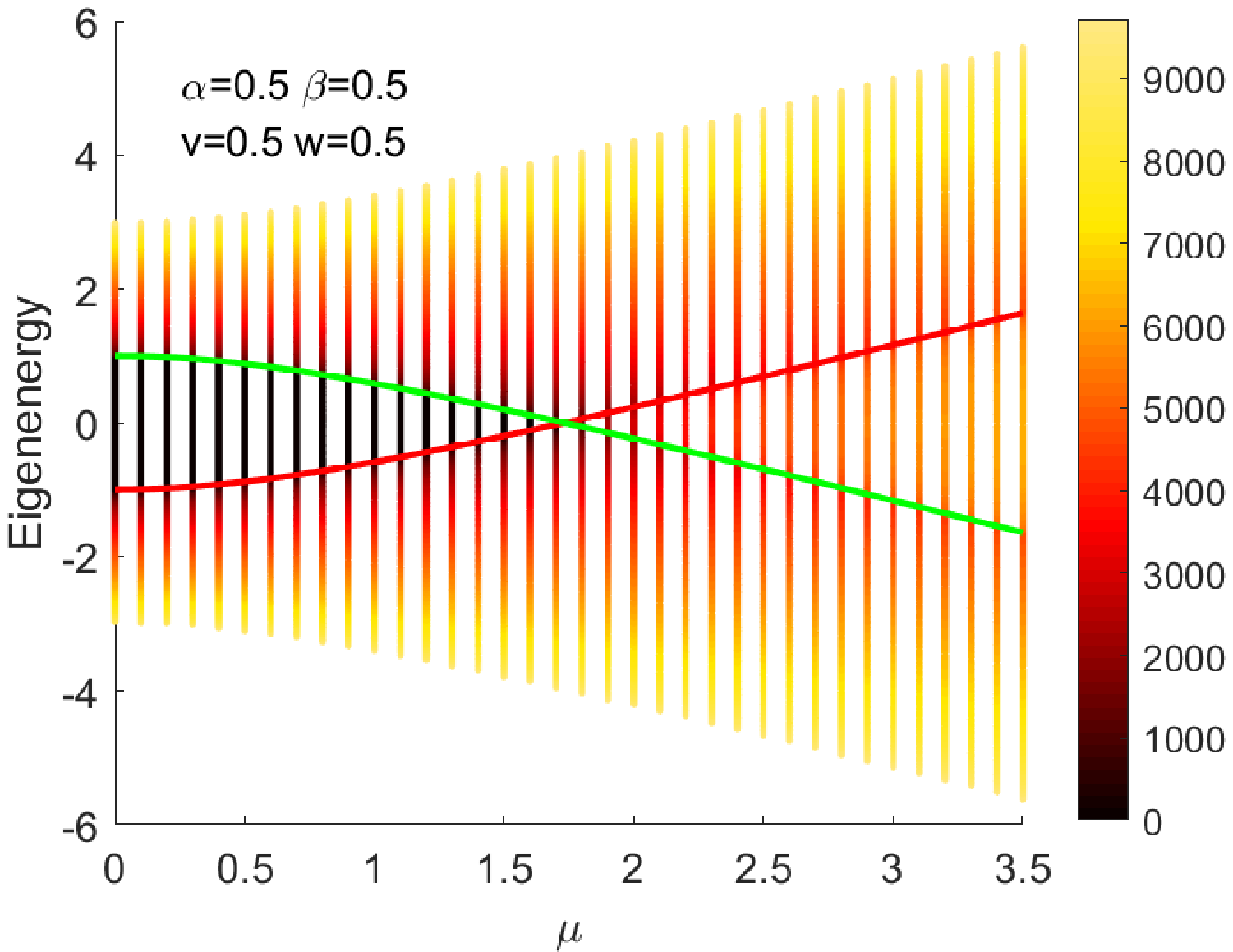}
  		\end{minipage}
  	}
  	\subfigure[]{
  		\begin{minipage}[t]{0.45\linewidth}
  			\centering
  			\includegraphics[width=\textwidth]{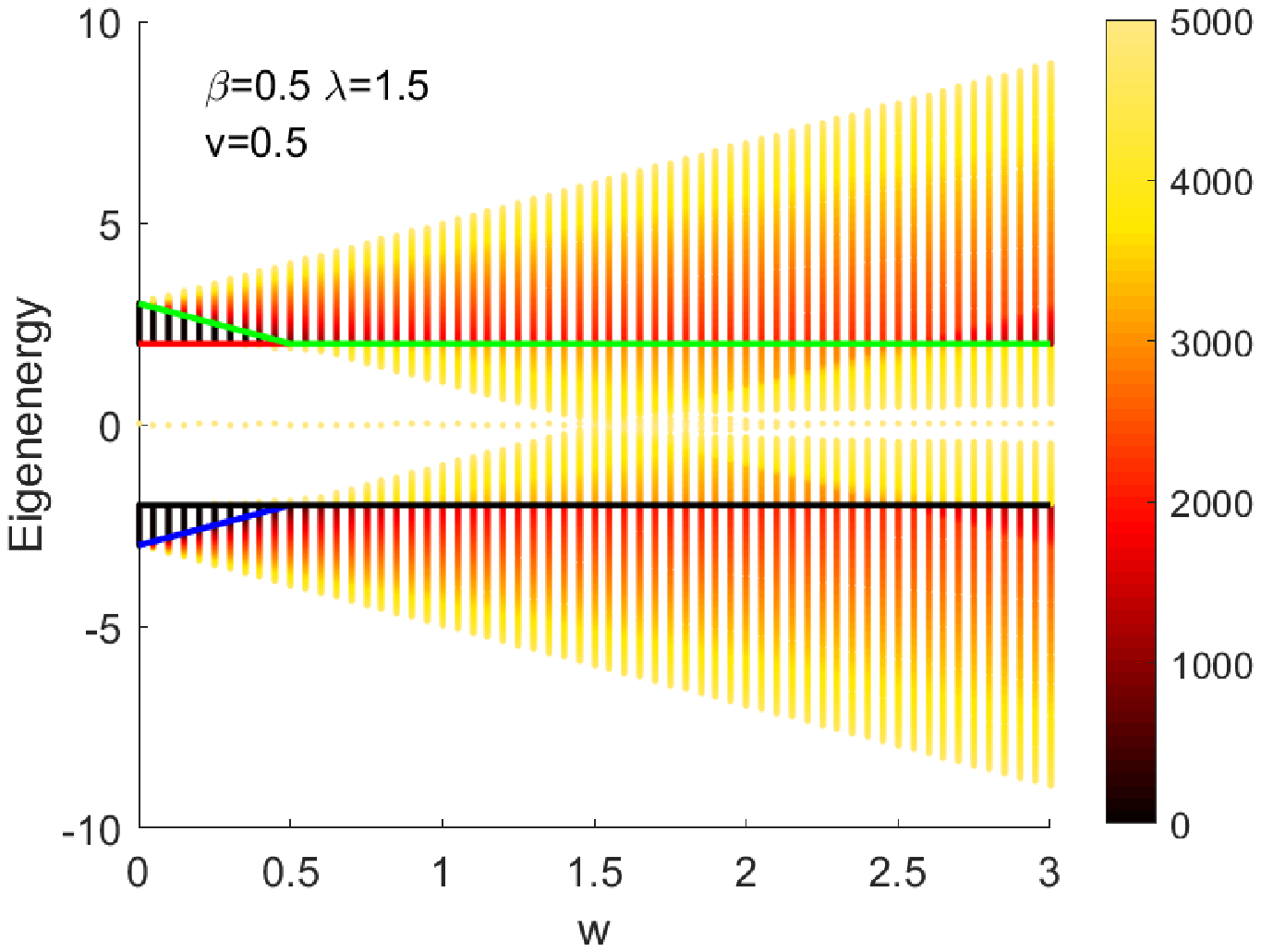}
  		\end{minipage}
  	}
  	\quad
  	\subfigure[]{
  		\begin{minipage}[t]{0.45\linewidth}
  			\centering
  			\includegraphics[width=\textwidth]{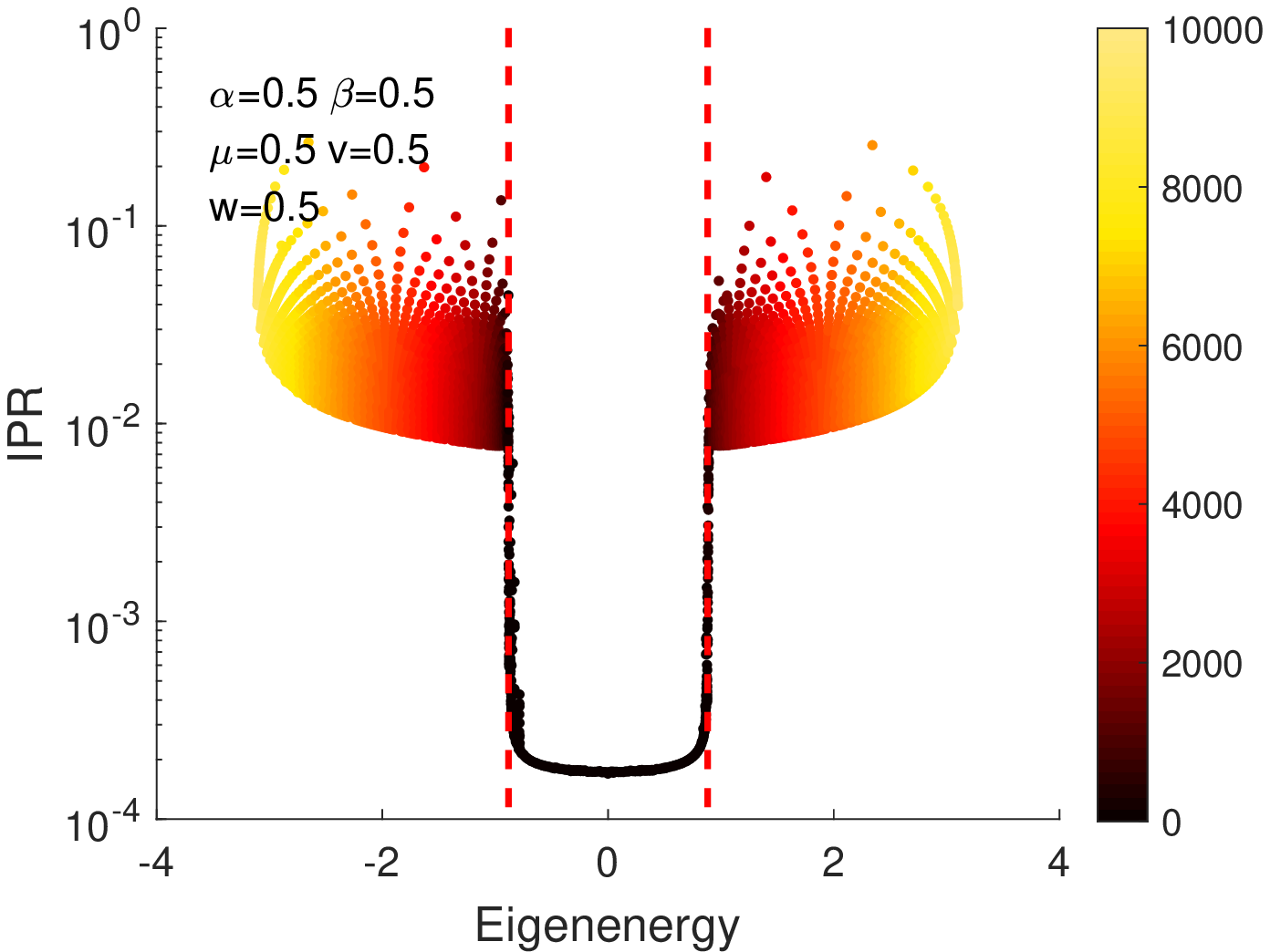}
  		\end{minipage}
  	}
  	\subfigure[]{
  		\begin{minipage}[t]{0.45\linewidth}
  			\centering
  			\includegraphics[width=\textwidth]{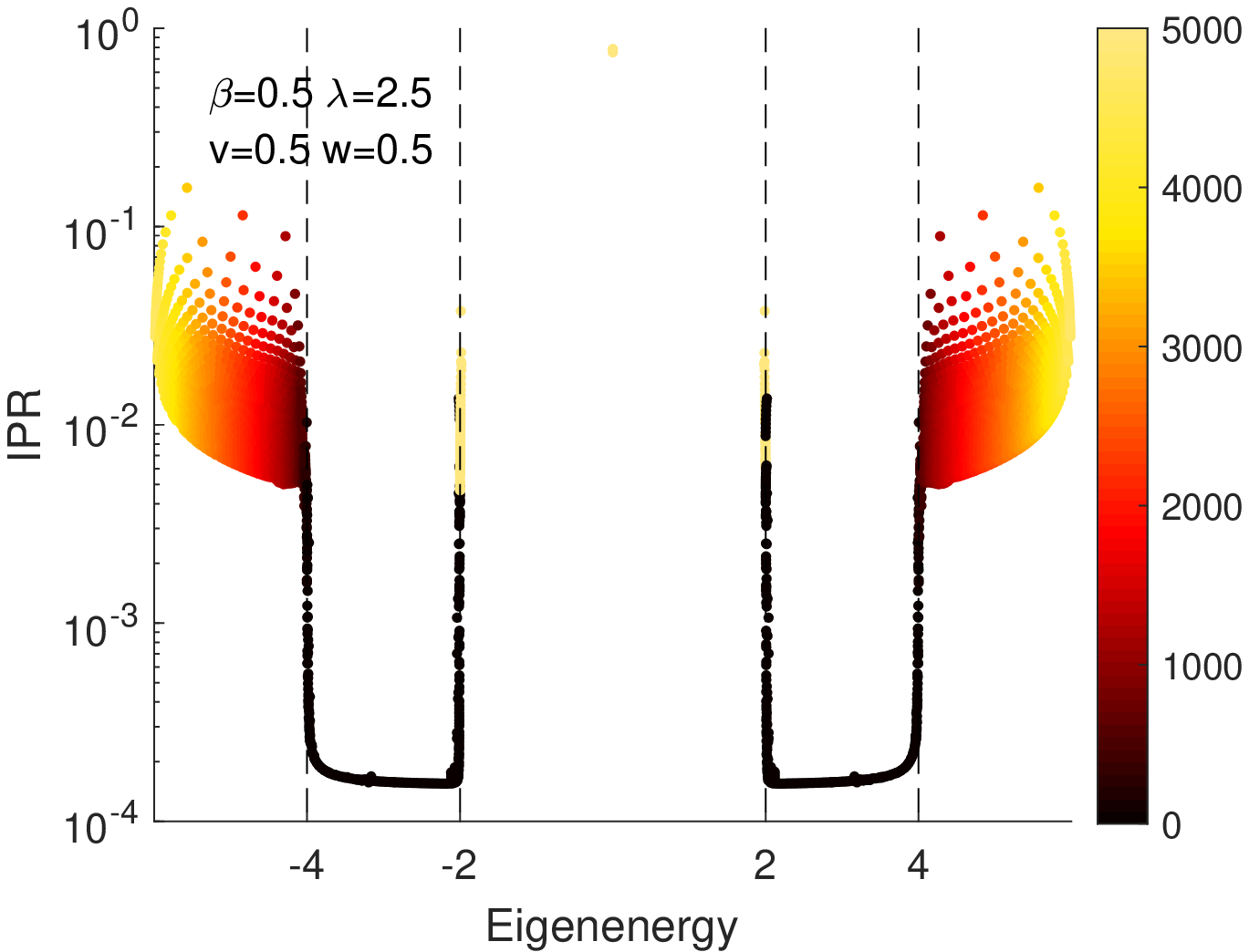}
  		\end{minipage}
  	}
  	\centering
\caption{(a) Eigenenergy of AA model Eq. (\ref{AAH}) as a function of $\mu$ with $\alpha=0.5$, $\beta=0.5$, $w=0.5$ and $v=0.5$, $\phi=\pi/2$.  (b) Eigenenergy of SSH model Eq. (\ref{SSH}) as a function of $mu$ with $\lambda=1.5$, $\beta=0.5$ and $v=0.5$. The colors in panels (a) and (b) represent the index of localization of each eigenstates. The red and green curves represent the mobility edges. (c) The IPR as a function of eigenenergy for $(w,\mu)=(0.5,0),\,(0.5,1),\,(0.5,2)$, other parameters are the same as (a). (c) The IPR as a function of eigenenergy for $(\lambda,\mu)=(1.5,0),\,(1.5,0.2),\,(1.5,0.4),\,(1.5,0.6)$, other parameters are the same as (b).  The colors in panels (c) and (d) represent the index of localization of each eigenstates.}
\label{fig-index}
  \end{figure}

The existence of mobility edge in the models with slowly varying quasi-periodic disorders is due to the coexistence of periodicity and disorders in the same model. In Eq.(\ref{dw}), the disorder term $w_i$ tends to be constant when lattice site $i$ is large enough. This represents the periodicity of the model which favors an extended wave-function. On the other hand, for small lattice site $i$, the term $w_i$ is quite random, which favors the localized wave-function. With these considerations, one would expect that how fast the $w_i$ term approaching a stable value is likely to affect the degree of localization. When $w_i$ approaches to a stable value at a rapid speed, the periodicity of the model is dominant, and most of the eigenstates of the model should be extended states. When the $w_i$ approaches to the stable value at a very slow speed, most of the eigenstates of the model should be localized states. If it is slow to a certain extent, the mobility edge used to separate the localized state from the extended state will disappear.

In the previous section, we have approximate the disordered model by an ensemble of periodic models $M_a$ for $a=1,\cdots,N$. The region of extended states of the disordered model is obtained by the overlap of the energy bands of all these periodic models $M_a$. This method actually suggest us a new index to represent the degree of localization at energy $E$. For a given energy $E$, we can find the number of models $M_a$ whose energy bands contains $E$. If this number equals the lattice site number $N$, we claim that $E$ belongs to the extended region. Therefore, this number actually reflect how extended this energy level $E$ is. The difference between $N$ and this number reflect how localized energy level $E$ is. With the above consideration, one can see that the index of localization $I(E)$ can be calculated as follows
\be
I(E)=N-\sum_{a=1}^{N}\int_0^{2\pi}dk\,\delta\Big(E-E_a(k)\Big)
\label{index}
\ee
For the AA model,  since the energy bands of model $M_a$ is $E(k)=\mu_a-2(t+w_a)\cos k$, we find
\be
I(E)=N-\sum_{a=1}^{N}\int_0^{2\pi}dk\,\delta\Big(E-[\mu_a-2(t+w_a)\cos k]\Big)
\ee

Now we show some numerical results about this new localization index. In Figure \ref{fig-index} (a) and (b), we plot the eigenenergy of AA model and SSH model as a function of $\mu$ and $w$ respectively. The parameters used in these calculations are the same as the one used in Figure \ref{fig-SSH} (a) and (b). The only difference is that the colors in Figure \ref{fig-index} represents the localization index we have just defined. Comparing to Figure \ref{fig-SSH}, one can see that the new index can also indicate the transition from localized states to extended states, just as IPR did. To make a more precise comparison, in Figure \ref{fig-SSH} (c) and (d), we plot IPR as a function of eigenerngy for the AA and SSH model for a fixed $\mu$ and $w$ respectively. In the same time, we use color to show the value of the localization index. One can see that the smaller values of IPR correspond to darker colors or smaller values of $I(E)$ and the larger values of IPR correspond to brighter colors or larger values of $I(E)$. This verifies that the new index can also indicate the degree of localization.

\section{Conclusion}

In this paper, we have presented a so called ``energy matching'' method to study the mobility edges of several 1D tight-binding models with slowly varying quasi-periodic disorders. In this method, the disordered model is approximated by an ensemble of periodic models. Then the region of extended states is determined by the overlap of the energy bands of all these periodic models. This method provides us a very efficient way to characterize the mobility edges in a model with complicated quasi-periodic disorders. Based on this method, we also proposed a new index which can help us to visualize the degree of localization of eigenstates.

\begin{acknowledgments}
This work is supported by the National Natural Science Foundation of China under Grant No. 11874272 and Science Specialty Program of Sichuan University under Grant No. 2020SCUNL210.
\end{acknowledgments}

\end{document}